\documentstyle[twocolumn,prl,aps,floats,epsf,lscape,grafik]{revtex}
\begin{document}
\draft
\preprint{}
\twocolumn[\hsize\textwidth\columnwidth\hsize\csname@twocolumnfalse\endcsname

\title{Disorder and Localization in the Lowest Landau Level}

\author{Z. Gedik and M. Bayindir}

\address{Department of Physics, Bilkent University, 
Bilkent 06533 Ankara, Turkey}

\date{\today}
\maketitle
\begin{abstract}

We study the localization property of a two-dimensional noninteracting
electron gas in the presence of randomly distributed short-range scatterers. 
We evaluate the  participation number of the eigenstates obtained by exact
diagonalization technique. At low impurity concentrations we obtain
self-averaged values showing that all states, except those exactly at the
Landau level, are localized with finite localization length. We conclude that
there is no universal localization exponent and at least at low
impurity concentrations localization length does not diverge.

\end{abstract}
\vskip1pc]

\narrowtext

There has been a long lasting interest in understanding the localization
problem in two-dimensional (2D) systems. According to scaling theory of
localization\cite{thouless,wegner,abrahams,sondhi}, all states in a 2D 
system are 
localized if a disordered potential is present. However, in the presence of a
strong perpendicular magnetic field, where the time reversal symmetry is
broken, extended states appear in the center of impurity-broadened Landau
bands\cite{aoki,ono}. If the scattering between Landau levels can be
neglected, these extended states exist only at a single energy\cite{chalker}.
The width of the quantized plateaus of the integer quantum Hall effect (QHE)
depends on the ratio of number of localized to extended states\cite{klitzing}. 

In analogy with the quantum critical phenomena and other localization
transitions, it has been proposed that localization length $\xi(E)$ diverges
as $E$ approaches the critical energy $E_c$, which is equal to Landau level
energy, so that
\begin{equation}
\xi(E)\propto|E-E_c|^{-\nu}
\label{diverge}
\end{equation}
where $\nu$ is the localization critical exponent\cite{trugman}. After the
initial calculations of Aoki and Ando\cite{ando1,aoki2,ando2}, several  
groups attempted to determine this critical exponent 
\cite{hikami,chalker2,pruisken,milnikov,huckestein,mieck,huo,janssen,lee,zhao,liu,ando3,huckestein2,matsuoka,varga}. 
Experimental results\cite{wei1,kawaji,koch,wei2,furlan} are generally in
good agreement with the calculated values. Various techniques have allowed the
computation of the exponent $\nu$, and they strongly suggest a universal value
close to $\nu=7/3$ for the lowest Landau level (LLL). However, in 
spite of a great deal of experimental evidence and numerical simulations in 
its favor, there is no rigorous derivation of power law divergence 
in the localization length. Furthermore, even if the power law divergence is 
true, it is not clear whether the localization critical exponent is 
universal, independent of impurity concentration or parameters of the 
disordered potential\cite{sarma}.

Recently, we developed a method for a particle in the LLL moving in an 
arbitrary potential\cite{gedik}. In this study we apply the  method, which
is basically an exact diagonalization technique, to a potential formed
by randomly distributed short-range scatterers. We concentrate on low impurity
concentrations where it is difficult to perform calculations by other methods
due to the presence of zero eigenvalues associated with the extended states at
the band center. At low enough concentrations, we obtain self-averaged values
where energy spectrum or localization property of eigenstates do not change
with increasing system size. Contrary to the widely accepted
view, localization length does not diverge at low impurity concentrations but
instead the maximum localization length grows exponentially with impurity
density. Extrapolation to less pure systems suggests that localization length
can become as large as the sizes of the samples used in QHE experiments which
explains the observed divergence in measurements. 

The Hamiltonian for a particle of mass $m$ and charge $q$, moving in 2D in
the presence of magnetic field ${\bf B}={\bf \nabla}\times{\bf A}$
perpendicular to the plane and potential $V$, is given by $H=H_{0}+V$ where 
\begin{equation}
H_{0}=\frac{1}{2m}({\bf p}-\frac{q}{c}{\bf A})^{2}~~~.
\end{equation}
Using the symmetric gauge 
${\bf A}=\frac{1}{2}{\bf B}\times{\bf r}$ and complex coordinates 
$z=X+iY=\sqrt{qB/2\hbar c}(x+iy)$ where ${\bf r}=(x,y)$, the unperturbed
Hamiltonian can be written as $H_0=\hbar\omega (a^\dagger a+1/2)$ where
$a^{\dagger}=-\partial/\partial z+z^*/2$. Since 
$[a, a^\dagger ] = 1$, the energy eigenvalues are given
by $E_n=\hbar\omega(n+1/2)$ where $\omega=qB/2mc$ ($q$ is 
assumed to be positive) and $n$= 0, 1, 2, ....
When the magnetic field is very high the particle
is confined into the LLL. This is a good 
approximation as long as the potential is small in comparison to 
Landau level splitting $\hbar\omega$. We are going to measure energies from
the LLL so that $E_n=0$. 

Now, let us consider the potential
\begin{equation}
V(z,z^*)=V_0\sum_{i} \delta(z-z_i)\delta(z^*-z_i^*)
\end{equation}
where $z_i$ denotes the position of the $i$th impurity in complex
coordinates defined above. 
According to our method\cite{gedik}, to find the nonzero eigenvalues, 
the matrix 
to be diagonalized is
\begin{equation}
\langle i|{\tilde V}|j\rangle=\frac{V_0}{\pi}
\exp({z_iz_j^*-|z_i|^2/2-|z_j|^2/2})~~~.
\end{equation}
Once ${\tilde V}$ is diagonalized, the eigenfunctions $\psi(z,z^*)$
of $V$ can be constructed from ${\tilde \psi}_i$  
\begin{equation}
\psi(z,z^*)=\sqrt{\frac{V_0}{\pi^2E}}\sum_i
\exp({zz_i^*-|z|^2/2-|z_i|^2/2}){\tilde \psi}_i~~~.
\label{trans}
\end{equation}

We distinguish between the extended and the localized states via participation
number $P$, which is the inverse of the mean fourth power of the
amplitude\cite{bell,wegner2}. Therefore, given a wave function $\psi$, 
$P$ is defined as
\begin{equation}
P=\frac{[\int|\psi({\bf r})|^2d{\bf r}]^2}{\int|\psi({\bf r})|^4d{\bf r}}.
\end{equation}
The participation number is a convenient quantity for distinguishing between 
localized and extended states since it takes a nonvanishing value for the 
former and becomes
infinite for the latter. If a state is localized within a $d-$dimensional
volume of average diameter $D$, $P$ behaves as $D^d$ irrespectively of the
system size. For a plane wave it depends on the system size as $L$ as $L^d$.
In general, extended states lead to an effective dimensionality $d^*$, smaller
than the real dimensionality $d$, which means that the states are not
space-filling\cite{schreiber}.

For ${\tilde \psi}$, participation number reduces 
to ${\tilde P}=1/\sum_i|{\tilde \psi}_i|^4$ provided that ${\tilde \psi}$ 
is normalized. For a state localized on single impurity ${\tilde P}=1$, while
for uniform distribution over $N_i$ impurities ${\tilde P}=N_i$. Therefore, we
can interpret ${\tilde P}$ as a measure of number of scatterers on which
${\tilde \psi}$ takes nonzero value. We note that, as can be seen from 
Eqn.~\ref{trans}, $\psi$ and ${\tilde \psi}$ have the same localization 
behavior, i.e. they are both extended or localized. Although we can evaluate 
corresponding $P$, we prefer to use ${\tilde P}$ to distinguish between 
localized and extended states. In this way we get rid of four-fold sums to 
be performed for evaluation of $P$ in terms of ${\tilde \psi}_i$. We note that 
for $f<1$, where $f$ is the number of impurities per flux quantum, there are
 $N_i(1/f-1)$ extended states at the center of the Landau band. However, the
 behavior given by Eqn. 1 has been  proposed for states with $E \neq E_c$ 
and our method filters the states with $E=E_c$.

\bfig{t}\ff{0.4}{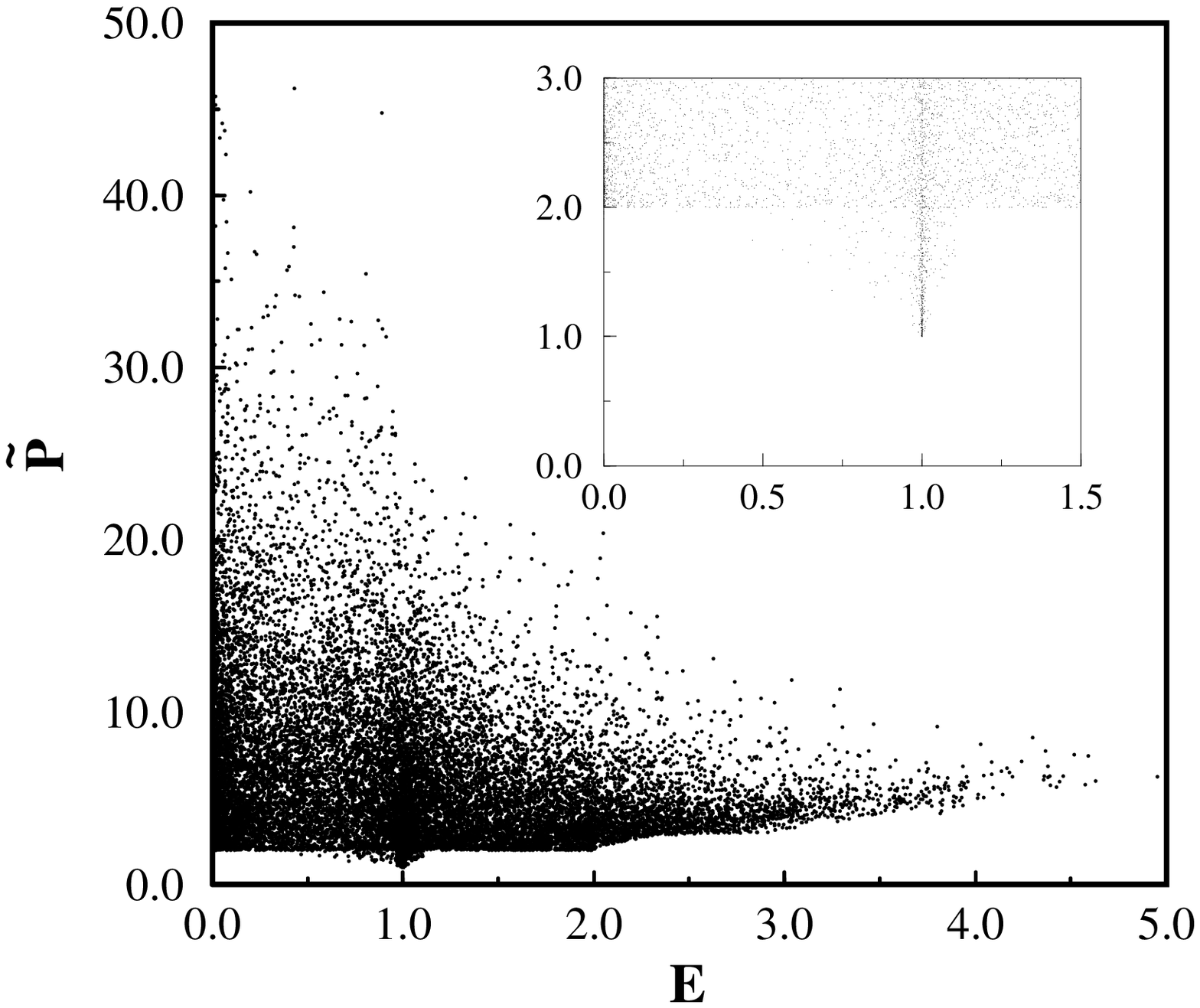}\efig{Participation number ${\tilde P}$ as a
 function of energy (measured in units of $V_0/\pi$) for $f=2/3$, i.e. two
 impurities per three flux quanta. There are $N_i=1250$ impurities and 16 
different distributions are used. The inset shows the details of the region
 around $E=1$.}{f1}

In Fig.~1, we plot participation number ${\tilde P}$ as a function of energy
measured in units of $V_0/\pi$. We keep the concentration of impurities
the same ($f=2/3$) and evaluate the energy eigenvalues and the eigenfunctions 
for different distributions of the scatterers. For this example, $N_i=1250$
and the number of configurations is 16 so that there are $2\times10^4$
points in the graph. Figure~1 shows that participation
number is direct measure of localization length. ${\tilde P}=1$ states, which
are localized on a single impurity, occur mainly at $E=1$ as we expect.
The inset shows the details of the region around $E=1$. Localization around a
purely repulsive scattering center is due to confinement into the LLL. All
other states involve at least two scatterers so that ${\tilde P}\ge 2$. We use
${\tilde P}$ to decide whether a state is extended or not as follows. Let us
consider $N_i$ impurities distributed in a square so that their concentration,
i.e. number of impurities per flux quantum, is $f$. If 
${\tilde P}>\sqrt{N_i}$, then ${\tilde \psi}$ is nonvanishing on
approximately $\sqrt{N_i}$ sites which means that it may extend from one side
of the square to its opposite. Therefore, we assume that ${\tilde \psi}$ is an
extended state. On the  other hand, if ${\tilde P}<\sqrt{N_i}$, then 
${\tilde \psi}$ has no chance to be extended. In this way we obtain the number
of extended states $N_e$, for a given system composed of $N_i$ impurities. 
Although ${\tilde P}>\sqrt{N_i}$, corresponding state can still be
localized. Therefore, $N_e$ is only an upper bound for the number of
extended states. Our definition of extendedness become exact if the sites at
which ${\tilde \psi}$ is nonvanishing form straight lines at least in one
direction. 

Figure~2 shows variation of $N_e$ with $N_i$ for different impurity
concentrations. Each point is obtained 
in such a way that number of different configurations times $N_i$
is $10^5$. 
Straight lines indicate that the two numbers  are related by
$N_e\propto N_i^y$ and the inset shows variation of $y$ with $f$. We note
that $y$ depends upon impurity concentration. If we assume that the maximum
localization length diverges as energy $E$ approaches the critical value $E_c$
(see Eqn.~\ref{diverge}), then it is easy to show that $\nu=1/2(1-y)$. The
widely accepted value $\nu=7/3$ corresponds to $y=11/14$ which we obtain at
$f=0.97$. At lower $f$ values, $\nu$ becomes lower. However, we must be
careful in comparing our results with the values in the literature since
most, in fact to our knowledge all, of the existing calculations have been
performed at impurity concentrations higher than unity. In addition to this,
our definition of extendedness is somewhat arbitrary. In general, the number
of extended states is less than $N_e$.  Finally, as we
are going to see from the next figure, divergence of the localization length
can merely be an artifact of finite size calculations.

\bfig{t}\ff{0.4}{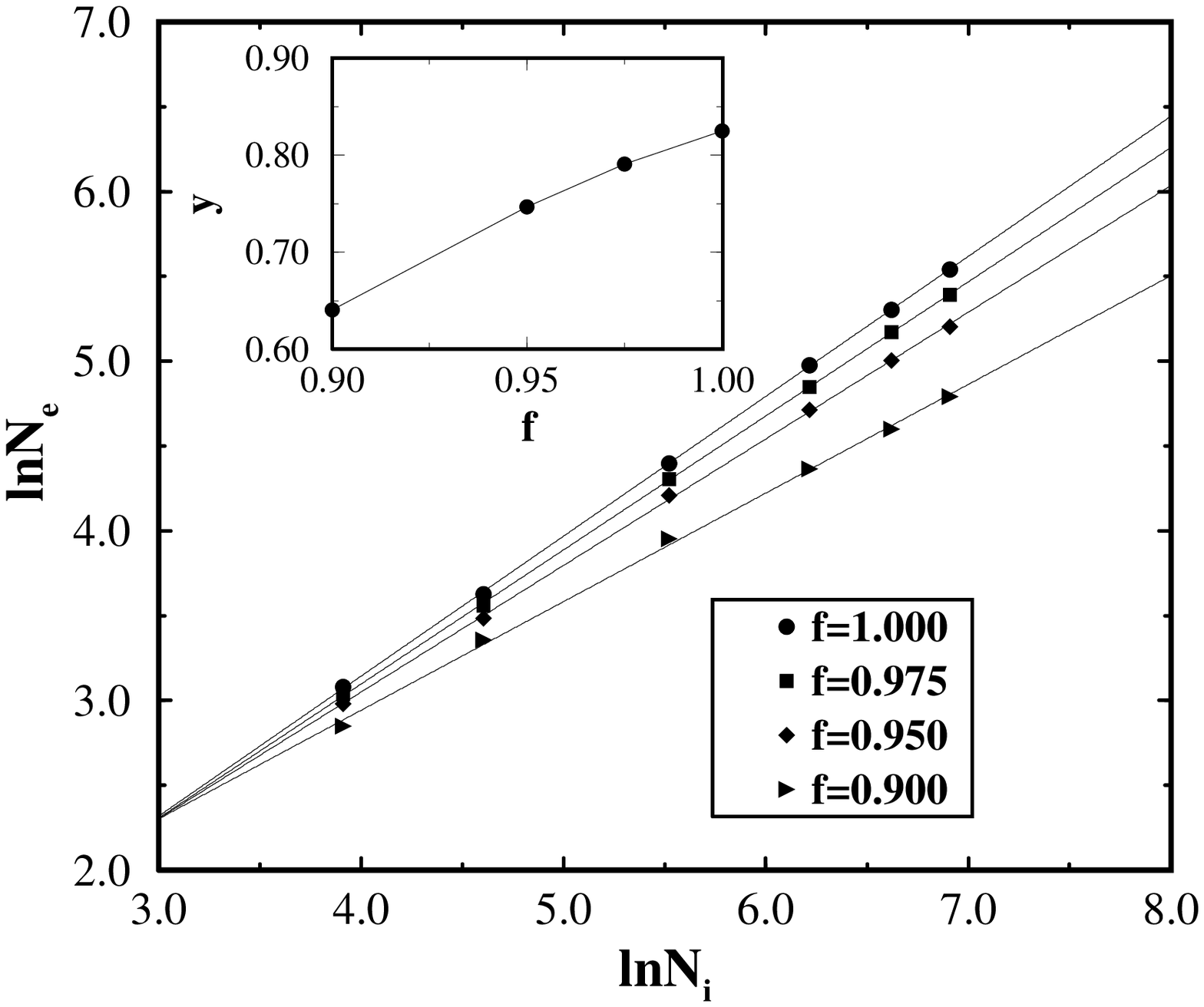}\efig{Number of extended states $N_e$ for a
 given number of impurities $N_i$. Linear fitting gives a relation
 $N_e\propto N_i^y$. Each point is obtained in such a way that the number
 of different configurations times $N_i$ is $10^5$. The inset shows the
 variation of $y$ with impurity  concentration $f$.}{f2}

As we go to more dilute systems and perform calculations in large enough 
systems a remarkable change occurs. As shown in Fig.~3, $N_e$ no more
increases with $N_i$ but instead decreases and vanishes at the end. For
$N_i>N_i^*$ there are no extended states. We define $N_i^*$ as the number of
impurities for which $N_e=1$. If we increase $N_i$ further, we do not get
any extended states. This result is independent of the arbitrariness of our
method of distinguishing between extended and localized states.

At this stage, it is not possible to say whether there
is no localization length divergence for
larger concentrations but Fig.~4 gives some idea about the system sizes to
be used if the same result holds in this denser region. In Fig.~4, we plot 
$N_i^*$ as a function of $f$. We observe that $N_i^*$ increases
exponentially which suggests that for
large enough sizes it may be possible to observe the same behavior even for
higher concentrations. If this is the case, then Fig. 2 is only an artifact 
of finiteness of the systems that we used in our calculations.

 We also performed calculations for the case with
equal number of attractive and repulsive point scatterers. We observe that
there is no qualitative change in our results as can be seen from the figure.
Beyond a certain size all states are localized and localization length does
not diverge.

\bfig{t}\ff{0.4}{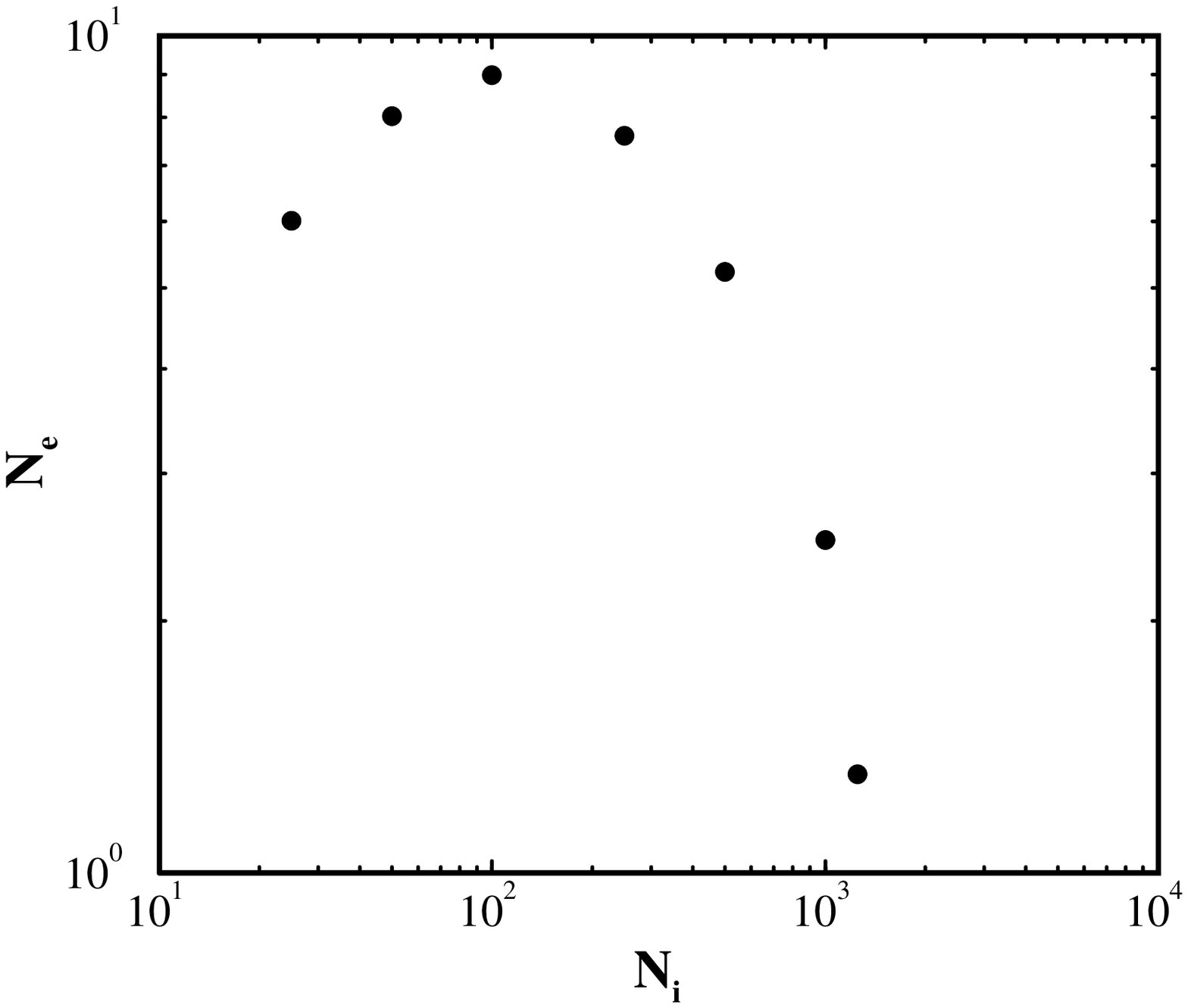}\efig{Number of extended states $N_e$ vs number 
of impurities $N_i$ for $f=2/3$. After $N_i=N_i^*=1598$ no extended states
 appear.}{f3}

It is known that for very clean samples QHE is not observed\cite{koch}.
Assuming that each impurity atom corresponds to a single $\delta$-function
potential, we evaluate the concentration $f$ 
but the resulting values (0.01-0.1) are too small. However,
it should be noted that in our model scatterers are short-ranged while in the
experiment opposite is true, i.e. disorder potential is smoothly varying in
comparison to the magnetic length. In
that case we can use a correlated disordered potential where several
$\delta$-functions are used to model the extended potential created by a
single impurity atom. 
Another point that should be taken into account before quantitative
comparison is that at low impurity concentrations electron-electron
interactions become important. 

\bfig{h}\ff{0.4}{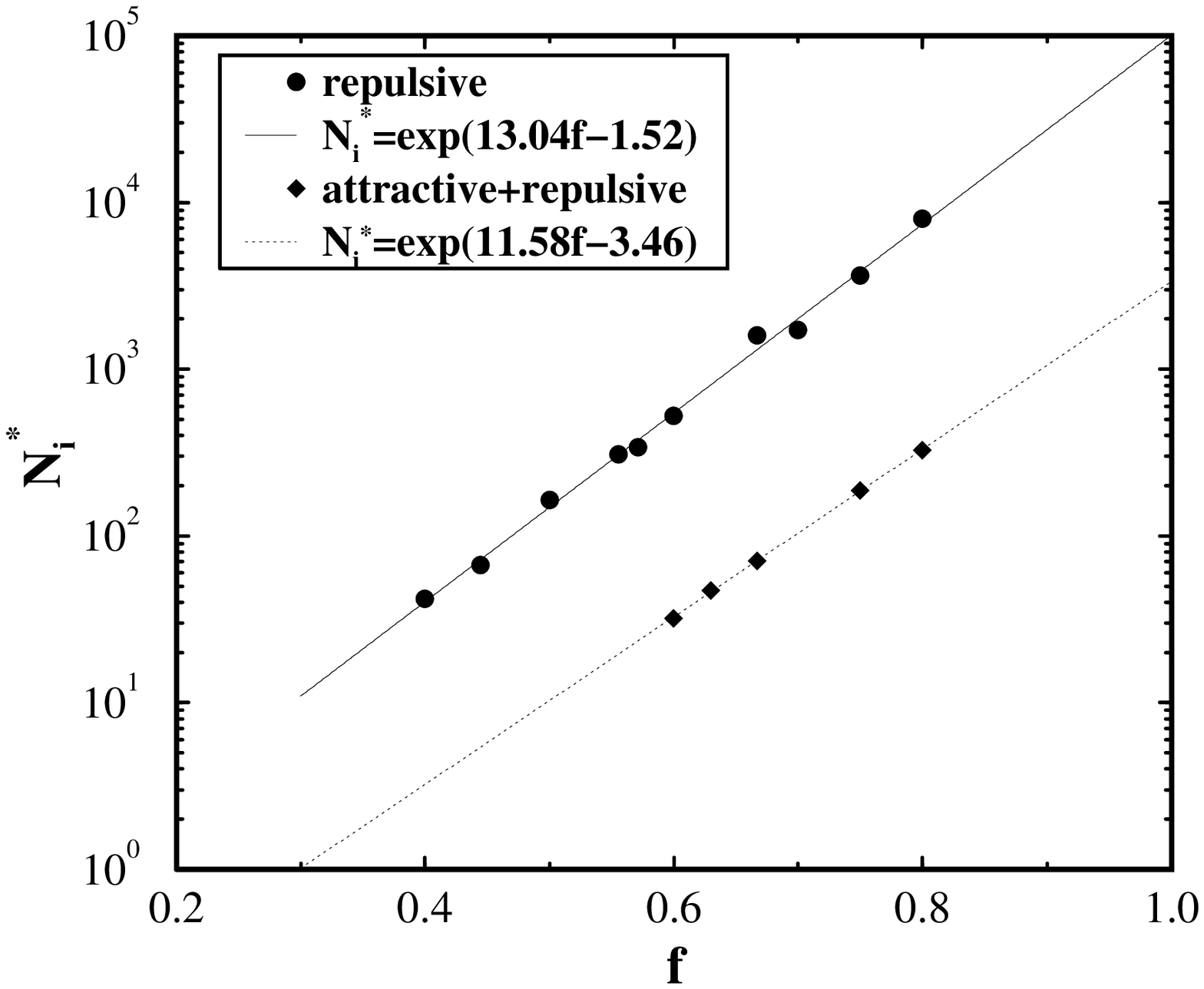}\efig{Variation of $N_i^*$ with $f$. The upper
 line is for pure repulsive case while the lower is obtained for the case 
with equal number of attractive and repulsive scattering centers.}{f4}

In conclusion, at low impurity concentrations we obtain
self-averaged values showing that all states, except those exactly at the
Landau level, are localized with finite localization length. We conclude that
the localization exponent is not universal and at least at low
impurity concentrations localization length does not diverge. Our results
suggest that the same behavior can be observed for higher concentrations.
If this is the case then there is no universal localization exponent, and 
in contrast to results of previous theoretical, numerical, and experimental studies, localization length does not diverge. On the other hand, if for higher concentrations there is a divergence, the transition  between the two regimes  is a very interesting problem to investigate. 

\acknowledgements{
This work was supported by the Scientific and Technical 
Research Council of Turkey (TUBITAK) under grant No. TBAG-1736. 
We gratefully acknowledge useful discussions with 
Prof. I.O. Kulik.}


\begin{references} 
\bibitem{thouless}D. J.~Thouless, Phys. Rep.  {\bf 13}, 93 (1974).
\bibitem{wegner}F.~Wegner, Z. Phys. B {\bf 25}, 327 (1976); {\em ibid} 
{\bf 35}, 207 (1979).
\bibitem{abrahams}E.~Abrahams, P.W.~Anderson, D.C.~Licciardello, and
T.V.~Ramakrishnan, Phys. Rev. Lett. {\bf 42}, 673 (1979).
\bibitem{sondhi}S. L.~Sondhi, S. M.~Girvin, J. P.~Carini, and D.~Shahar, 
Rev. Mod. Phys. {\bf 69}, 315 (1997).
\bibitem{aoki}H.~Aoki and T.~Ando, Solid State Commun. {\bf 38}, 1079 (1981).
\bibitem{ono}Y.~Ono, J. Phys. Soc. Jpn. {\bf 51}, 2055 (1982). 
\bibitem{chalker}J.T.~Chalker, J. Phys. C {\bf 20}, L493 (1987). 
\bibitem{klitzing}K.~von Klitzing, G.~Dorda, and M.~Pepper, Phys. Rev. Lett. 
{\bf 45}, 494 (1980).   
\bibitem{trugman}S.A.~Trugman, Phys. Rev. B {\bf 27}, 7539 (1983).
\bibitem{ando1}T.~Ando, J. Phys. Soc. Jpn. {\bf 52}, 1740 (1983). 
\bibitem{aoki2}H.~Aoki and T.~Ando, Phys. Rev. Lett. {\bf 54}, 831 (1985).   
\bibitem{ando2}T.~Ando and H.~Aoki, J. Phys. Soc. Jpn. {\bf 54}, 2238 (1985).
\bibitem{hikami}S.~Hikami, Prog. Theor. Phys.  {\bf 76}, 1210 (1986).
\bibitem{chalker2}J.T.~Chalker and P.D.~Coddington, J. Phys. C {\bf 21},
2665 (1988).
\bibitem{pruisken}A.M.M.~Pruisken, Phys. Rev. Lett. {\bf 61}, 1297 (1988).
\bibitem{milnikov}G.V.~Milnikov and I.M.~Sokolov, JETP Lett. {\bf 48}, 536 
(1988).
\bibitem{huckestein}B.~Huckestein and B.~Kramer, Phys. Rev. Lett. {\bf 64},
1437 (1990).
\bibitem{mieck}B.~Mieck, Europhys. Lett. {\bf 13}, 453 (1990). 
\bibitem{huo}Y.~Huo and R.N.~Bhatt, Phys. Rev. Lett. {\bf 68}, 1375 (1992). 
\bibitem{janssen}U.~Fastenrath, M.~Janssen, and W.~Pook, Physica A {\bf 191}, 
401 (1992). 
\bibitem{lee}D.H.~Lee, Z.~Wang, and S.~Kivelson, Phys. Rev. Lett. {\bf 70},
4130 (1993).
\bibitem{zhao}H.L.~Zhao and S.~Feng, Phys. Rev. Lett. {\bf 70}, 4134 (1993).
\bibitem{liu}D.~Liu and S.~Das Sarma, Phys. Rev. B {\bf 49} 2677 (1994).
\bibitem{ando3}T.~Ando, Phys. Rev. B {\bf 49} 4679 (1994).
\bibitem{huckestein2}B.~Huckestein, Rev. Mod. Phys. {\bf 67}, 357 (1995).
\bibitem{matsuoka}H.~Matsuoka, Phys. Rev. B {\bf 55} R7327 (1997).
\bibitem{varga}I.~Varga {\em et al.}, cond-mat/9710036.
\bibitem{wei1}H.P.~Wei, D. C.~Tsui, M. A.~Paalanen, and A. M. M.~Pruisken,  
Phys. Rev. Lett. {\bf 61}, 1294 (1988).  
\bibitem{kawaji}J.~Wakabayashi, M.~Yamane, and S.~Kawaji, J. Phys. Soc. 
Jpn. {\bf 58}, 1903 (1989).
\bibitem{koch}S.~Koch, R.J.~Haug, K.~von Klitzing, and K.~Ploog, 
Phys. Rev. B {\bf 43}, 6828 (1991); {\em ibid} {\bf 46}, 1596 (1992).
Phys. Rev. Lett. {\bf 67}, 883 (1991).   
\bibitem{wei2}H.P.~Wei, L.W.~Engel, and D.C.~Tsui, Phys. Rev. B {\bf 50}, 
14609 (1994).  
\bibitem{furlan}M.~Furlan, Phys. Rev. B {\bf 57}, 14818 (1998). 
\bibitem{sarma} S. Das Sarma,  in {\em Perspectives in Quantum Hall Effects}, 
edited by S.D.~Sarma and A.~Pinczuk, (Wiley, New York, 1997).
\bibitem{gedik}Z.~Gedik and M.~Bayindir, Phys. Rev. B {\bf 56}, 12088 
(1997).   
\bibitem{bell}R.J.~Bell and P.~Dean, Discuss. Faraday Soc. {\bf 50}, 55
(1970).
\bibitem{wegner2}F.~Wegner, Z. Phys. B {\bf 36}, 209 (1980).
\bibitem{schreiber}M.~Schreiber, Phys. Rev. B {\bf 31}, 6146 (1985).    
\end{references}
\end{document}